# Nuclear Magnetic Resonance Study of Ultrananocrystalline Diamonds

A. M. Panich[1*], A. I. Shames[1], H.-M. Vieth[2], E. Ōsawa[3], M. Takahashi[3], A. Ya. Vul'[4]

[1] Department of Physics, Ben-Gurion University of the Negev, Beer Sheva 84105, Israel,

[2] Institute for Experimental Physics, Free University Berlin, Arnimallee 14, D-14195 Berlin, Germany

[3] NanoCarbon Research Institute, Ltd., 5-4-19 Kashiwa-no-ha, Kashiwa, Chiba 277-0882, Japan

[4] Ioffe Physico-Technical Institute, 26 Polytechnicheskaya, St.Petersburg 194021, Russia,

Short title: NMR Study of Ultrananocrystalline Diamonds

* Corresponding author. Fax: +972-8-6472903, Tel.: +972-8-6472458,

E-mail address: pan@bgu.ac.il (A.M. Panich)

**Abstract**

We report on a nuclear magnetic resonance (NMR) study of ultrananocrystalline diamond (UNCD) materials produced by detonation technique. Analysis of the $^{13}$C and $^{1}$H NMR spectra, spin-spin and spin-lattice relaxation times in purified UNCD samples is presented. Our measurements show that UNCD particles consist of a diamond core that is partially covered by a *sp$^2$*-carbon fullerene-like shell. The uncovered part of outer diamond surface comprises a number of hydrocarbon groups that saturate the dangling bonds. Our findings are discussed along with recent calculations of the UNCD structure. Significant increase in the spin-lattice relaxation rate (in comparison with that of natural diamond), as well as stretched exponential character of the magnetization recovery, are attributed to the interaction of nuclear spins with paramagnetic centers which are likely fabrication-driven dangling bonds with unpaired electrons. We show that these centers are located mainly at the interface between the diamond core and shell.

# 1. Introduction

One of the effective synthetic routes for the mass production of nanodiamond materials, making them to be commercially available, is the detonation technique [1]. Detonation of TNT-hexogen oxygen-imbalanced composition in inert atmosphere like $CO_2$ or water produces 4-5% of soot, which leaves about one half of its weight after oxidation with hot nitric acid. The unoxidized residue can be worked up to grey fine powder, which comprises complex agglomerates of tiny nanodiamond particles. Intensive sonication of aqueous suspension of agglomerates followed by stirred-media milling with micron-sized zirconia beads broke up core agglutinates to give a transparent colloidal solution of primary particles of nanodiamond crystals having an average diameter of 4.3±0.3 nm as determined by X-ray diffraction [2] and confirmed by TEM observations [2-4] and dynamic light scattering (DLS) measurements [5].

Single cubic crystals of nanodiamond in this size-range, 2-5 nm, also occur in CVD diamond films being developed by Gruen and his coworkers in the past decade, who called these particles as "ultrananocrystalline diamond" (UNCD) [6]. We will adopt this naming for the primary particles of detonation nanodiamond studied here. Recently published *ab initio* and semiempirical molecular orbital calculations [7-11] show a variety of UNCD structures. Using *ab initio* calculations, Raty et al. [7-9] have proposed an explanation for the ultradispersity of diamond on the nanoscale based on simple thermodynamics arguments. They showed that, depending on temperature and pressure of the hydrogen and carbon gases used in the diamond growth process, diamond would grow into nanoparticles of about 3 nm with non-hydrogenated, fullerene-like surface. The authors believe that such a structure is energetically more favorable than a hydrogenated surface. However, calculations of Barnard [10] demonstrate that the surfaces of dehydrogenated nanodiamonds are structurally unstable, while the surface hydrogenation may induce some stability. At the same time, our high-resolution transmission electron microscopy (HRTEM) experiments [2-4] reveal that the diamond core of the UNCD particles is covered by a shell, which is presumably built of $sp^2$-hybridized carbon atoms. The objective of the present investigation is a detailed $^{13}$C and $^{1}$H NMR study of the structure and crystalline quality of a series of UNCD samples prepared by the detonation technique. We note that NMR is a very useful tool in studying the structural features and determination of different allotropic forms in such compounds since the position of the NMR signal of each nucleus depends on the nature of chemical bonding.

## 2. Experimental details

We studied a series of detonation nanodiamond samples at various stages of purification. Raw detonation soot (I) was prepared at Diamond Centre, St. Petersburg, and used as received. Sample II is a commercial 'Diamond Nano-Powder' produced at Gansu Lingyun Nano-Material Co. Ltd, Lanzhou, China, in September 2003 by oxidizing raw detonation soot with hot nitric acid [1]. The light gray Gansu powder consists of complex aggregates with no significant distribution of free primary particles [5], but was analyzed in this work as received.

Sample III is an aqueous dispersion of near-primary particles obtained by stirred-media milling of commercial nanodiamond powder (Sample II) with ceramics microbeads. In a typical run, 250 ml of 10% suspension of Sample II in water was circulated through 160 ml zirconia-plated stainless steel mill (Ultra Apex Mill UAM-015 manufactured by Kotobuki Eng & Manufact. Co. Ltd, Tokyo) loaded with yttrium-stabilized zirconia beads having average diameter of 30 μm (manufactured by Netsuren Co., Japan) to 60% of the effective milling volume. The content of mill was stirred with rotating horizontal bars attached to a vertical axis, all plated with zirconia, at a periphery speed of 8 m/s for 40 min. The circulating system was dynamically closed with a pair of mechanical seal at the top and bottom of mill and purged with nitrogen to avoid oxidation with air. The progress of milling was monitored by taking out aliquots from the circulating suspension, diluted to *ca* 0.1% concentration and subjected to DLS measurements of particle size. At the end of milling period, 90 wt% of particles had a size-distribution at 8.4±0.8 nm, corresponding to dimers of primary particles. Milling was stopped at this point in order to suppress contamination with zirconia at a minimum. Analysis of the solid residue of aliquot by Inductively Coupled Plasma Optical Emission Spectroscopy revealed the presence of 0.35% of zirconia.

Sample IV is loosely aggregated powder prepared by evaporating water from colloidal solution of disintegrated nanodiamond, which can be re-dispersed in water by applying intensive sonication. In the present work, a new batch of commercial nanodiamond from Gansu Lingyun Nano-Material Co. Ltd., prepared on August 2005 in China, was subjected to stirred-media milling under similar conditions as used to prepare Sample III to obtain 10% colloidal solution containing more than 90% of near-primary particles of nanodiamond having average diameter of 8.3±1.5 nm. When water was removed from the freshly prepared colloidal solution using a rotary evaporator in a water bath heated at 60 ºC, drying was stopped while a small amount of water still remained. Whereas the optimum amount of water to be retained in the powder has not been determined, in this drying process 2.9% of water was left unevaporated. The moisture content was determined by thoroughly drying a small portion of the moist aggregate powder at 120 ºC for 2 h under a vacuum of 10 Pa and weighing the completely dried residue.

All NMR spectra have been measured at room temperature. Static $^{13}$C and $^1$H NMR spectra, spin-lattice and spin-spin relaxation times have been measured in the applied magnetic field $B_0 =$ 8.0196 T at resonance frequencies 85.85 MHz and 341.41 MHz, respectively. The pulse lengths were 3.9 and 1.4 µs for $^{13}$C and $^1$H NMR. The spin-lattice relaxation time $T_1(^{13}C)$ of the diamond core was measured using standard partially relaxed Fourier transform method [12]. In this method, the partially relaxed free induction decay (FID) after the reading pulse is Fourier transformed to yield a partially relaxed spectrum. Thus, from the spectrum as a function of the time between pulses, the spin-lattice relaxation time of the intensive peak, assigned to the diamond core, was obtained. At that, the mixing with $sp^2$-signal is negligibly small. $^{13}$C magic angle spinning NMR spectra were recorded at 150.8 MHz ($B_0$ = 14.09 T) using either $^1$H-$^{13}$C cross-polarization or high power $^1$H decoupling (HPDEC) techniques. $^{13}$C chemical shifts $\delta$ are given relative to tetramethylsilane (TMS).

## 3. Results

### 3.1 Static $^{13}$C NMR spectra

Static $^{13}$C NMR spectrum of the detonation soot (Fig. 1, insert) is a superposition of two lines attributed to the diamond particles and carbon soot. The latter is significantly reduced as a result of purification. The spectra of purified UNCD samples (Fig. 1) consist of a strong signal showing the line width $\Delta\nu \approx 2.1$ kHz and chemical shift $\delta$ ($^{13}$C) = 35 ± 2 ppm, and a weak broad tail at the high frequency domain. The chemical shift of the narrow signal is characteristic of bulk diamond [13-15], thus this signal is definitely attributed to the $sp^3$-carbons belonging to the diamond core. The broad signal is assigned both to the shell that covers the UNCD particles and to the modified diamond surface. It is likely a superposition of several overlapping signals that are unresolved due to the (i) $^{13}$C chemical shielding anisotropy and (ii) $^{13}$C-$^1$H dipole-dipole coupling that occurs in the hydrocarbon groups. We note that the spectra of all purified samples are rather similar despite the difference in the treatment procedure.

### 3.2 $^{13}$C MAS NMR spectra

High-resolution $^{13}$C spectra of the purified UNCD samples have been measured using the magic angle spinning (MAS) technique that effectively averages out the anisotropic nuclear interactions and yields narrow NMR lines. Let us first discuss $^1$H-$^{13}$C cross-polarization (CP) MAS measurements. This technique selectively brings out carbon atoms with a neighboring hydrogen nucleus (i.e., C atoms bound to H directly or sometimes even through one or two C–C bonds). The CPMAS spectra of the samples II and IV (Fig. 2) exhibit several peaks related to the hydrogenated carbons. The peak at $\delta \approx 44$ ppm is assigned to the signals of the $CH_2$ and CH groups, the chemical shifts of which vary in the range of 20-55 ppm [13]. The peak at $\delta \approx 73$ ppm is assigned to the C-OH group. Besides that, the peak at 34-35 ppm is assigned to tetrahedral carbons of the diamond phase. We note that the aforementioned hydrocarbon groups may result from a reaction between the carbon shell or diamond surface with a hydrogen-containing precursor during detonation, as well as with nitric acid used for the sample purification. A component at $\delta \approx 119$ ppm, corresponding to aromatic carbons without any hydrogen atoms nearby, is suppressed in these $^1H$-$^{13}C$ CPMAS measurements. However, a narrow $sp^2$-carbon signal (at $\delta=111$ ppm) is well seen in the "common" MAS spectra measured at thermal equilibrium without cross-polarization (Fig. 3). This mode detects all carbon atoms including "hydrogen-free" carbons. The aforementioned signal is attributed to a fullerene-like shell that covers the diamond core. Our data are in accordance with the interpretation of the HRTEM measurements [4], which show occurrence of an onion-like carbon shell with the separation between layers being 0.35 nm. The small width of this signal ($\sim 3$ ppm) reflects the decent perfection of the carbon shell. The ratio of $sp^3$ to $sp^2$ carbons in our samples varies from ~6.4 in sample II to ~8.8 in sample IV. This result will be discussed below. We note that a broad $^{13}C$ signal at ~120 ppm, corresponding to $sp^2$- carbons, was observed by Donnet et al. [16] in some of their nanodiamond samples, while in the other samples, differing by treatment procedure, such a signal was not detected. Alam [17] and Belobrov et al. [18] did not report $sp^2$-carbons signal in nanodiamonds at all.

### 3.3 $^{13}C$ spin-spin and spin-lattice relaxation measurements

$^{13}C$ spin-spin relaxation time ($T_2$) measurements (Fig. 4) show that the magnetization decay in all samples is well described by a superposition of at least two exponentials:

$$M(t) = M_1(0) \times \exp(-(t/T_{21})) + M_2(0) \times \exp(-(t/T_{22})) \qquad (1)$$

It is clearly seen from the semi-logarithmic plot of the magnetization decay (Fig. 4, bottom). This fact strongly supports existence of at least two different carbon species. Here, the longer $T_2$ is assigned to the narrower NMR line belonging to the diamond core, while the shorter $T_2$ is assigned to the broader line belonging to the shell and surface carbons. $T_2$ values of different samples are given in Table 1.

Measurements of the $^{13}$C nuclear spin-lattice relaxation time $T_1$ of the δ=35 ppm carbon resonance (i.e. signal of the diamond core) show that the magnetization recovery in all purified UNCD samples is well described by a stretched exponent

$$M(t) = M(0)\{1 - \exp[-(t/T_1)^\alpha]\}, \qquad (2)$$

where M(0) is the equilibrium magnetization. This is clearly seen in Fig. 5 that shows the variation of the magnetization as a function of recovery time to the power of 0.57. The values of $T_1$ and parameters α for different samples are given in Table 1. The anomalous reduction in the $^{13}$C spin-lattice relaxation time from several hours in natural diamond [14, 19-21] to hundred milliseconds in UNCD, as well as the stretched exponential character of the magnetization recovery are attributed to the interaction of nuclear spins with some paramagnetic centers. The theory of such relaxation [22-24] shows that the parameter α=1 is typical for the case of rapid nuclear spin diffusion, while in the regime of vanishing spin-diffusion it is equal to D/6 or (D + d)/6 for uniform and non-uniform distributions of paramagnetic centers and nuclei, respectively [23, 24]. Here D is the space dimensionality of the sample (for diamond, D = 3), and d is the dimensionality of the nuclear magnetization space; high magnetic field NMR experiment yields d = 1 [23, 24].

There are two reasons why rapid nuclear spin diffusion in the compounds under study is unlikely. First, because of the low natural abundance (1.1%) of the $^{13}$C isotope (the only carbon isotope having nuclear spin) the compound under study is a dilute system with respect to nuclear spins. Therefore spin diffusion, driven by the dipole-dipole interaction causing mutual flips of adjacent nuclear spins, is expected to be significantly weaker than in spin-concentrated systems. Second, it is known that nuclear spin diffusion is quenched inside the diffusion barrier radius $b_0$ around a paramagnetic center, where the difference in local magnetic fields at adjacent nuclei, which arises due to the unpaired electron spin, is high enough to prevent the flip-flop process. In this region, a process of direct relaxation dominates. Using the approach described by Khutsishvili [25], $b_0$ in diamond was estimated to be 1.4 nm at T=295 K and an external magnetic field $B_0$=8 T. Thus $b_0$ is comparable with the size of diamond core. Reynhardt and High [26] calculated that spin diffusion in diamond is practically quenched if the density of the paramagnetic centers is above $N_e \approx 8 \times 10^{18}$ spin/cm$^3$. In our UNCD compounds, $N_e \approx 2.5 \times 10^{19}$ spin/cm$^3$ was determined by EPR [4, 27]. Therefore we conclude that the regime of vanishing spin-diffusion is a rather good approximation for the samples under study. Assuming a uniform and a non-uniform distribution of the paramagnetic centers yield $\alpha_{theor}$= 0.5 and 0.66, respectively. Our experimental data fall between these two values (Table 1). Suggesting that both distributions are reasonable in the UNCD samples, one can notice satisfactory agreement between the theory and experiment. The $T_1$'s of the diamond core for all samples, including the one that was de-aggregated in water, are of the same order of magnitude and rather close to each other (Table 1). The estimated relaxation times of the shell/surface signal are by the order of magnitude shorter than those of the diamond core (Table 1), revealing stronger interaction of the outer carbon nuclei with the paramagnetic centers. This fact indicates that the paramagnetic centers are likely located near the surface rather than in the diamond core, which may be reflected in a non-uniform distribution. Here, the diamond core carbons are in the average field of exchange-coupled surface-located spins and could indicate a uniform distribution. Altogether this is evidence that the observed parameter α reflects some intermediate value between these two limiting cases.

### 3.4 $^1$H NMR spectra and relaxation

The $^1$H NMR spectra of all UNCD samples consist of two lines (Fig. 6). The broad component with the line width $\Delta\nu\approx31$ kHz is attributed to closely set rigid hydrogen atoms, likely related to rigid $CH_2$ groups, while the narrow component showing $\Delta\nu\approx2.6$ kHz is assigned to C-H and C-OH groups at the surface.

Spin-spin relaxation time ($T_2$) measurements show that the magnetization decay in the samples under study is well described by a superposition of two exponentials corresponding to two resonance lines observed in the experiment. It is clearly seen from the semi-logarithmic plot of the magnetization decay (Fig. 7). This fact confirms existence of two hydrogen species. Here, the longer $T_2$ is assigned to the narrower NMR line belonging to C-H and C-OH groups at the surface, while the shorter $T_2$ is assigned to the broader line belonging to the methylene groups. The $T_2$ values of the different samples are given in Table 2. $^1$H spin-lattice relaxation measurements show bi-exponential magnetization recovery; the corresponding values are given in Table 2.

### 4. Discussion

As it was shown above, UNCD particles comprise three components: nanodiamond core, fullerene-like carbon shell that covers this core, and some hydrocarbon groups. Such structure does not conflict much with the conclusion of Barnard [10] that the surfaces of dehydrogenated nanodiamonds are structurally unstable, while the surface hydrogenation may induce some stability. However, the Barnard's model does not include an onion-like shell that likely is a result of partial graphitization of diamond during the sample fabrication. Such diamond-graphite transformation at high temperature is well established [28]. The first traces of onion-like carbon were observed after UNCD annealing at 1600 K [29]. On the other hand, the nanodiamond temperature may rise up to 3270 K during explosion. At the milling process, based on our estimation, temperature can reach 1770 K on the shell due to collisions [30], and, as the result, graphite-like $sp^2$ carbon may be formed on the surface of 4.3 nm UNCD particle.

Furthermore, as we noticed above, the ratio of $sp^3$ to $sp^2$ carbons in our samples varies from ~6.4 in sample II to ~8.8 in sample IV. Calculations [31, 1] show that spherical diamond particle having an outer diameter of 4.3 nm consists of 7425 [31] or 7342 [1] carbon atoms. A monolayer fullerene-like shell that is necessary for complete covering of such a cluster would have around 2360 carbons. Such architecture should show a ratio of $sp^3$ to $sp^2$ carbon around 3.13, which is smaller than that determined in our experiment. It means that in our samples the outer diamond surface is only partially covered by such a shell, leaving some bare plots. We believe that these plots are targets for attack by the functional groups and are finally hydrogenated in some places. Such explanation partially reconciles the theoretical models of Raty et al [7-9] and Barnard [10] and suits well our NMR findings. A schematic illustration of UNCD particle with different surface groups and reconstructions was designed by Shenderova et al [32].

Let us now discuss the location of the paramagnetic centers in the UNCD particles. These centers were shown to come from the fabrication-driven dangling bonds with unpaired electron [4, 27]. Such bonds, being placed on the outer diamond surface, are chemically active, causing a reaction with hydrogen-containing precursor during detonation, as well as with hydrogen ions of nitric acid used for the sample purification. It yields UNCD particles with partially hydrogenated surfaces, where most of dangling bonds are saturated. Thus the outer UNCD surface should comprise rather small amount of the paramagnetic centers. At the same time, the broken bonds placed at the interface between the shell and diamond core, are inaccessible for the aforementioned saturation and remain paramagnetic. Such conclusion is supported by our spin-lattice relaxation measurements mentioned above and by the EPR measurements [4, 27]. Furthermore, EPR studies [4, 27] show a 10% increase of the actual amount of detectable paramagnetic centers under pumping. This fact was interpreted in the frames of strong exchange interaction of unpaired electrons with paramagnetic oxygen molecules located in close vicinity of the paramagnetic centers. Such interaction causes significant broadening of the EPR lines of these paramagnetic centers, thus reducing their contribution to the observed signal intensity obtained by the signal integration within the limited field range. It was concluded that 90% of the observed paramagnetic centers (broken bonds) are likely located at the interface between the diamond core and its shell. In such a case, they do not interact with absorbed oxygen. At the same time, 10% of the paramagnetic centers are placed at the outer shell surface and interact with oxygen, giving rise to the decrease of the EPR signal. This interpretation correlates with our model derived from the NMR data.

## 5. Summary

Our measurements show that UNCD particles consist of three parts. Central part is a diamond core built from $sp^3$-carbons. This core is partially covered by a fullerene-like shell of $sp^2$-carbons. The carbon atoms, located at the uncovered spots of the outer surface, are bound to hydrogen and oxygen atoms, producing a variety of functional hydrocarbon groups that saturate the dangling bonds. Such architecture stabilizes the nanodiamond structure. The inner part of the shell adjacent to the diamond core comprises a larger amount of dangling bonds than the diamond core.

**Acknowledgement**

Financial support by New Energy Development Organization of Japan (grant # 04IT4) is acknowledged.

**Figure captions**

Fig. 1. Static $^{13}$C NMR spectra of the detonation soot (insert) and purified UNCD samples II-IV.

Fig. 2. $^{1}$H-$^{13}$C CPMAS NMR spectra of the purified samples II and IV.

Fig. 3. HPDEC $^{13}$C MAS NMR spectra of the purified samples II and IV.

Fig. 4. $^{13}$C Hahn echo decay in the sample II ($T_2$ measurements) on the linear (top) and semi-logarithmic (bottom) scales. Dashed and solid lines on the top figure show single and two-exponential fit, respectively.

Fig. 5. $^{13}$C magnetization recovery ($T_1$ measurements) of sample III on a semi-logarithmic scale.

Fig. 6. $^{1}$H NMR spectrum of the sample II.

Fig. 7. $^{1}$H spin echo decay ($T_2$ measurements) of the sample II on a semi-logarithmic scale.

Table 1. $^{13}$C spin-lattice ($T_1$) and spin-spin ($T_2$) relaxation times and parameter α from Eq.2. In the last row, parameters of the purified sample studied by us in Ref. [4] are also shown.

| Compound | $T_1$, ms (diamond core) | $T_1$, ms (shell) | α | $T_2$, μs (diamond core) | $T_2$, μs (shell) |
|---|---|---|---|---|---|
| II | 455 ± 13 | 36 ±12 | 0.589 | 1820 ± 107 | 308 ± 40 |
| III | 415 ± 19 | 22 ±15 | 0.572 | 1758 ± 171 | 383 ± 73 |
| IV | 377 ± 10 | 34 ±11 | 0.583 | 1826 ± 93 | 301 ± 35 |
| UNCD sample studied in Ref.[4] | 266 ± 13 | 24 ±11 | 0.594 | 1676 ± 85 | 328 ± 43 |

Table 2. $^1$H spin-lattice ($T_1$) and spin-spin ($T_2$) relaxation times.

| Compound | $T_{11}$, ms | $T_{12}$, ms | α | $T_{21}$, μs | $T_{22}$, μs |
|---|---|---|---|---|---|
| II | 24.4 ± 0.3 | 1.34 ± 0.08 | 0.87 | 41 ± 3 | 367 ± 60 |
| IV | 24.8 ± 1.1 | 1.1 ± 0.02 | 0.85 | 44 ± 5 | 217 ± 50 |

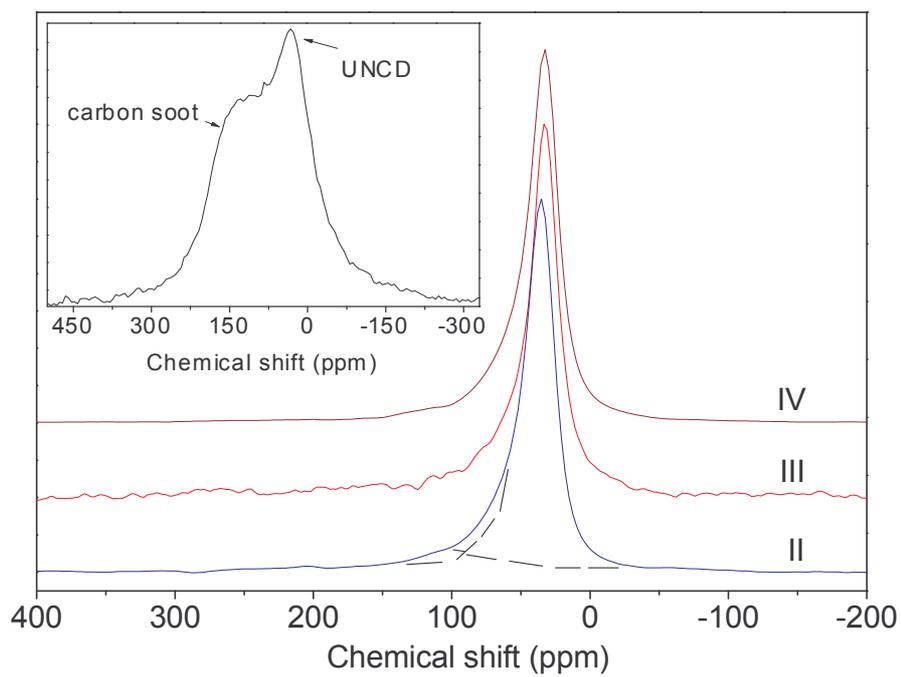

Fig. 1. Static $^{13}$C NMR spectra of the detonation soot (insert) and purified UNCD samples II-IV.

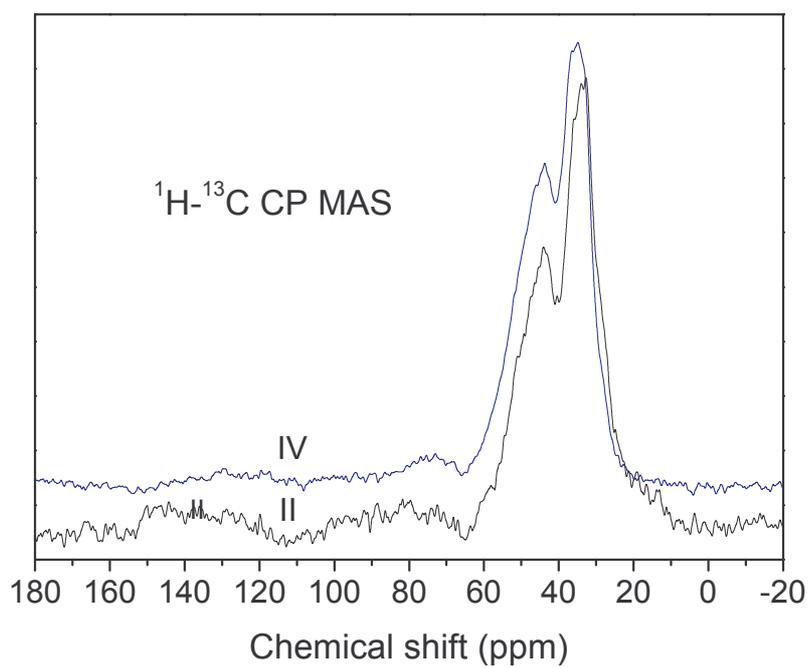

Fig. 2. $^1$H-$^{13}$C CPMAS NMR spectra of the purified samples II and IV.

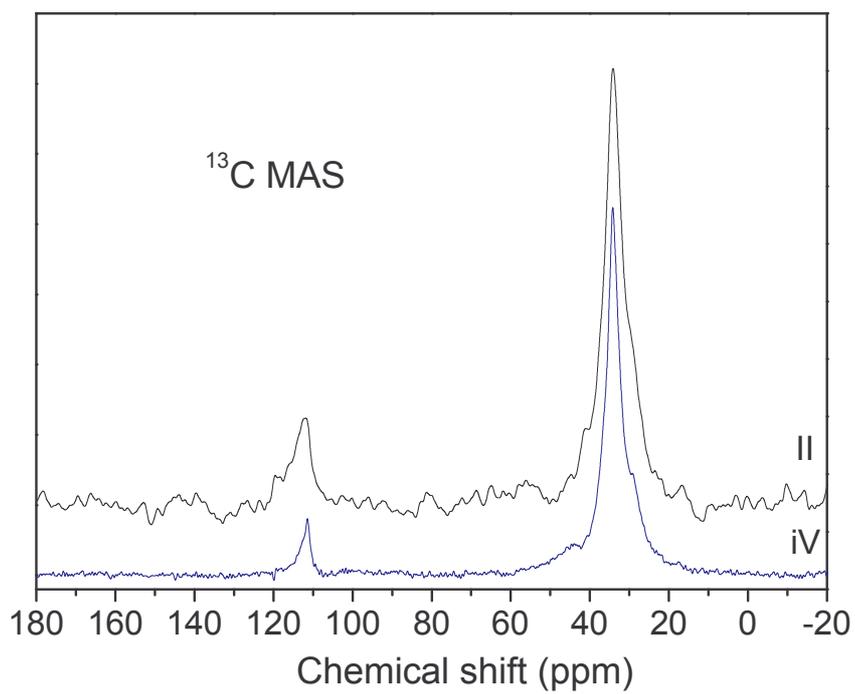

Fig. 3. HPDEC $^{13}$C MAS NMR spectra of the purified samples II and IV.

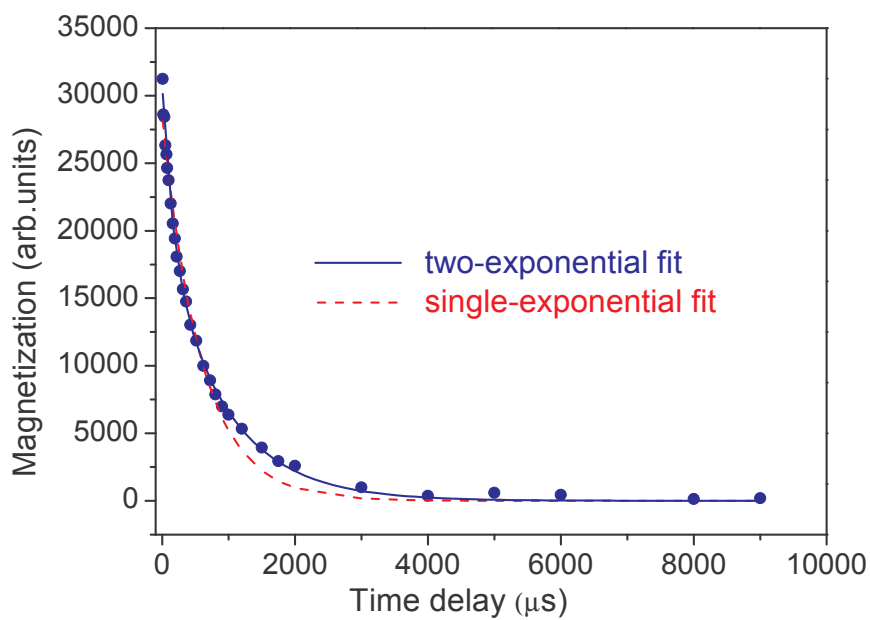

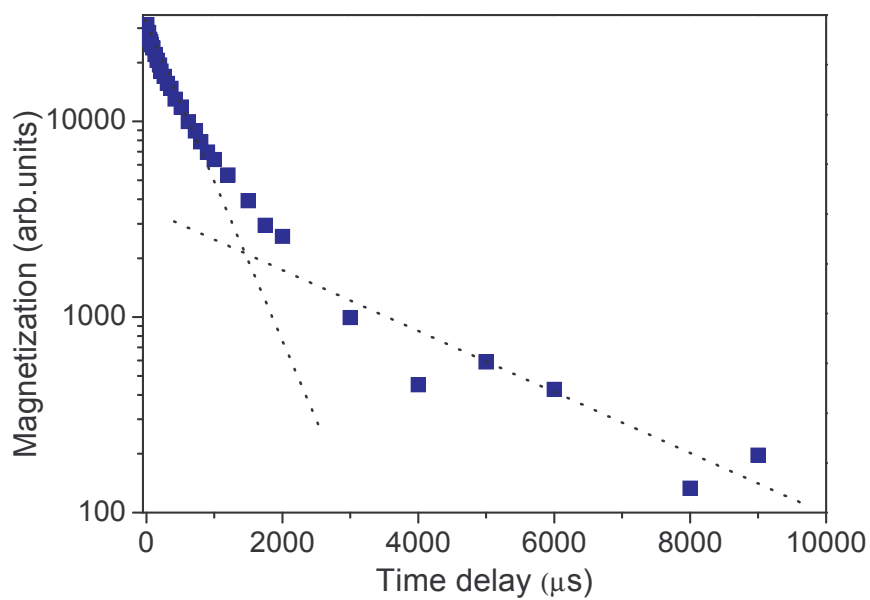

Fig. 4. $^{13}$C Hahn echo decay in the sample II ($T_2$ measurements) on the linear (top) and semi-logarithmic (bottom) scales. Dashed and solid lines on the top figure show single and two-exponential fit, respectively.

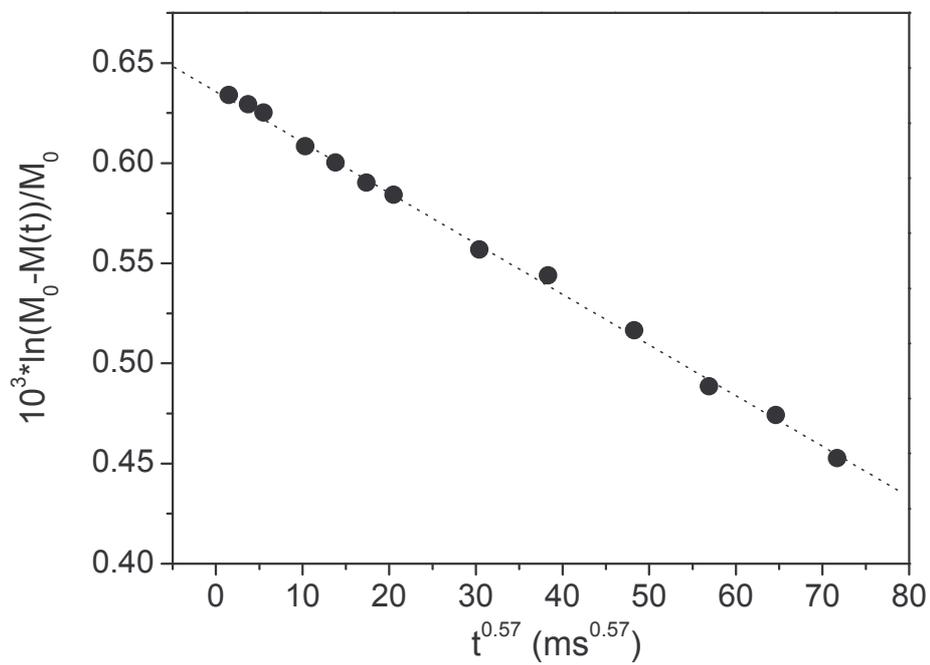

Fig. 5. $^{13}$C magnetization recovery ($T_1$ measurements) of sample III on a semi-logarithmic scale.

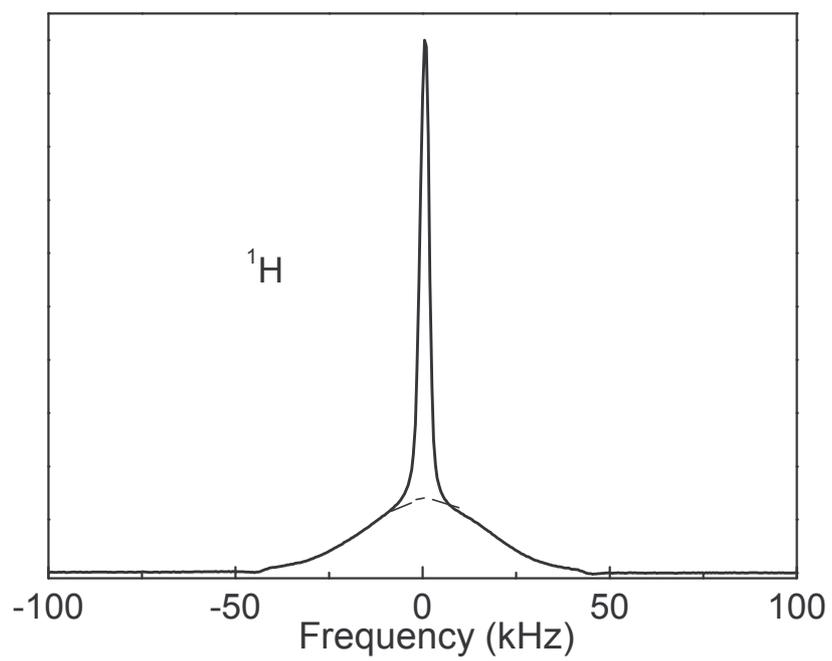

Fig. 6. $^1$H NMR spectrum of the sample II.

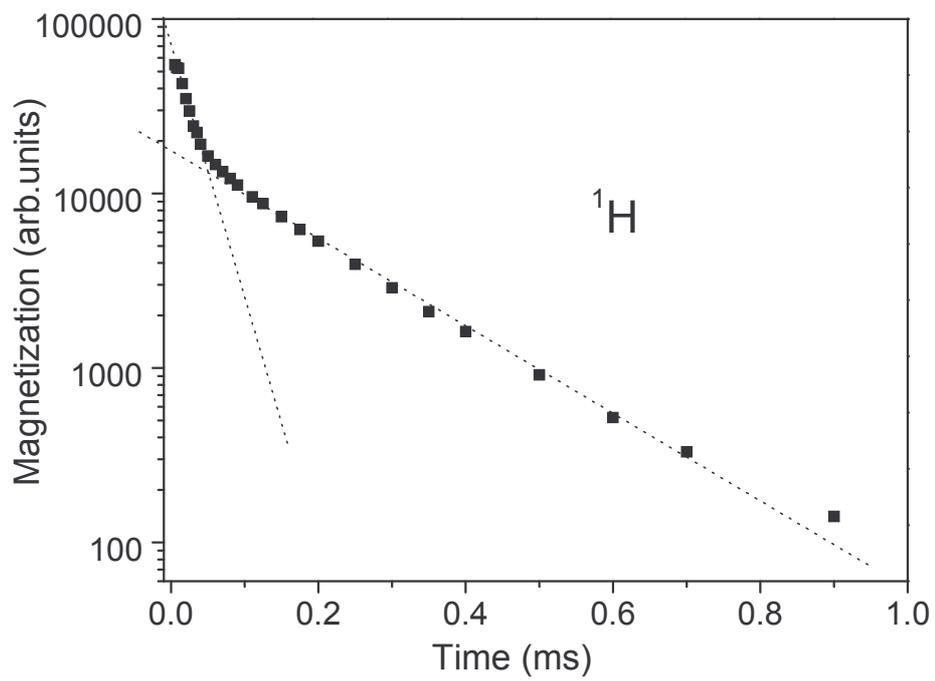

Fig. 7. $^1$H spin echo decay ($T_2$ measurements) of the sample II on a semi-logarithmic scale.